\documentclass[12pt,preprint]{aastex}

\usepackage{graphicx}
\usepackage{epsfig}
\usepackage{subfigure}
\usepackage{amsthm,amsmath,amssymb}
\usepackage{mathrsfs}
\usepackage[displaymath]{lineno}
\usepackage{booktabs}
\usepackage{multirow}

\shorttitle{Jitter radiation for TeV emission of GRB 221009A}

\shortauthors{Mao et al.}

\begin{document}

\title{Jitter Mechanism as a Kind of Coherent Radiation: Constrained by the GRB 221009A Emission at 18 TeV}
\author{Jirong Mao$^{1,2,3}$, and Jiancheng Wang$^{1,2,3}$
}
\affil{$^{1}$
Yunnan Observatories, Chinese Academy of Sciences, 650011 Kunming, Yunnan Province, People's Republic of China; jirongmao@mail.ynao.ac.cn
\\
       $^{2}$
Center for Astronomical Mega-Science, Chinese Academy of Sciences, 20A Datun Road, Chaoyang District, 100012 Beijing, People's Republic of China; 
\\
       $^{3}$
Key Laboratory for the Structure and Evolution of Celestial Objects, Chinese Academy of Sciences, 650011 Kunming, People's Republic of China; \\
       }

\begin{abstract}
The emission of gamma-ray burst (GRB) 221009A at 18 TeV has been detected by the large high-altitude air shower observatory (LHAASO). We suggest jitter radiation as a possible explanation for
the TeV emission for this energetic GRB. In our scenario, the radiation field is linked to the perturbation field,
and the perturbation field is dominated by kinetic turbulence. Kinetic turbulence takes a vital role in both magnetic field
generation and particle acceleration. The jitter radiation can reach the TeV energy band when we consider
either electron cooling or Landau damping. 
We further suggest that the jitter radiation in the very high-energy band is coherent emission.
Our modeling results can be constrained by the observational results of GRB 221009A in the TeV energy band.
This radiation mechanism is expected to have wide applications in the high-energy astrophysical research field. 
\end{abstract}


\section{Introduction}
Radiation mechanisms are one of the key points in astrophysical research. In the framework of the lepton model,
synchrotron radiation is a traditional mechanism to explain the emission of high-energy objects.
Alternatively, the so-called jitter radiation has been considered by some researchers in recent years.
Meanwhile, radiation mechanisms are associated with particle acceleration, and sometimes plasma effects are involved in both radiation and acceleration
processes.

High-energy phenomena of some objects have been searched by high-energy telescopes and detectors.
For example, gamma-ray bursts (GRBs) have been detected in the TeV energy band in recent
years \citep{abdalla19,magic19a,magic19b}. Nowadays, observational signatures challenge GRB radiation models \citep{hess21}. 
In particular, the large high-altitude air shower observatory (LHAASO) detected the emission of GRB 221009A that can reach the frequency of 18 TeV
\citep{lhaaso22}. This energetic GRB was also detected by Fermi-LAT, and a photon of 99 GeV was identified 240 s after the trigger \citep{omodei22}.   

When we consider the synchrotron mechanism to produce GRB prompt emission, it is found that electrons do not cool sufficiently \citep{ghisellini20}.    
A large of dissipation energy with $\epsilon_e>0.1$ is injected in a fraction of only less than 1\% electrons \citep{daigne11}.
While GRB emission in the GeV-TeV energy band is
reproduced by inverse Compton (IC) scattering, which usually has efficient cooling.
Moreover, high-energy emission of GRB is treated in the afterglow case. However,
the TeV emission reproduced by the IC mechanism with the multiwavelength data fitting shows a derivation from common afterglow model assumptions \citep{derishev21}.
Furthermore, the magnetic field strength should be in the condition of $\epsilon_B<10^{-3}$ for IC scattering to reproduce the TeV emission. For example, GRB 221009A provides the number of $\epsilon_B\sim 10^{-6}-10^{-7}$ when the external shock model is considered \citep{gonzalez22}.
The assumption, which is the energy equipartition given to electrons and magnetic field, is broken down.
Some GRBs show magnetically dominated features. In such cases, the magnetic energy should be effectively dissipated \citep{fraija20,gill20}.
In order to further investigate the GRB physics, one may consider some other radiation processes coupled with microhydrodynamics and plasma effects.  

We utilize jitter radiation to explain the TeV emission of GRB 221009A in this paper.
A preliminary exploration of jitter radiation toward GRB GeV-TeV emission was given by \citet{mao11} and \citet{mao21}.  
Here, we suggest this radiation mechanism to reproduce the emission of GRB 221009A in the TeV energy band.
In our scenario, the jitter radiation is in the perturbation theory case. The radiation field is related to the perturbation field,
and the kinetic turbulence is suggested. Then, Landau damping in the kinetic turbulence is introduced to constrain the maximum energy of the jitter radiation.  
We further suggest that the jitter radiation in our scenario can be treated as a kind of coherent emission.

We present a mini review in Section 2.1 to describe the jitter radiation that has been studied. It is important to stress the major physics of
this radiation mechanism. After the model description in Section 2.2, we calculate the constraints to the TeV emission of GRBs. The results for GRB 221009A is
given in Section 2.4. A few issues related to radiation and particle acceleration are discussed in Section 3, and the conclusion is drawn in Section 4.

\section{Jitter Radiation: A kind of Coherent Emission in the Perturbation Theory}

\subsection{Mini Review}
Synchrotron radiation is the relativistic electron radiating in the dipolar and large-scale magnetic field. The radiation frequency
is limited to be about 160 MeV. Through IC scattering, synchrotron photons can be scattered by the external
electrons or by the electrons themselves that produce the synchrotron photons. Traditionally, synchrotron and IC mechanisms have been
used to explain the emissions of high-energy objects above the MeV-GeV energy bands.

Jitter radiation is the relativistic electron radiating in the random and small-scale magnetic field.
Here, we only focus on the perturbation theory. 
\cite{medvedev99} suggested the Weibel instability to generate the magnetic field that has the length scale of the plasma skin depth. 
\citet{medvedev00} and \citet{medvedev06} presented the general presentation of the jitter radiation. In the works, the acceleration term
to produce the jitter radiation is put by the form of the magnetic field in a direct way. The jitter radiation in general was taken as a kind of incoherent radiation. 
\citet{fleishman06} derived a simplified equation to calculate the jitter flux using the perturbation theory. \citet{kelner13} revisited the jitter radiation
and pointed out that the radiation frequency might be extended toward the high-energy bands.  

We present the differences between jitter radiation and synchrotron radiation. First, the magnetic field length scale can be examined by the
Larmor radius, which is written as $r_L=\gamma m_ec^2/eB$, where $\gamma$ is the electron Lorentz factor. The corresponding Larmor frequency is $\omega_L=eB/\gamma m_ec$.
If the length scale of the magnetic field is much larger than the Larmor radius, we traditionally
consider synchrotron radiation. If the length scale of the magnetic field is roughly comparable to the Larmor radius, we may use jitter
radiation. Second, in the cases of synchrotron radiation, 
different electrons move in the magnetic field with different pitch angles and have different radiation cones.
The relativistic electrons lose energy and have radiation.
In the case of the jitter mechanism, in our scenario, as the radiation field is coupled with a certain perturbation field,
plasma turbulence, for example, is introduced as an origin of the perturbation.
The orbit of each electron moving in the random magnetic field is similar to a straight line, and each radiation cone is roughly along the line of sight.
Third, an electron with a certain energy has a synchrotron radiation in a very
narrow radiation frequency range,
while an electron has the jitter radiation in a wide radiation frequency range (see Equation (1) in Section 2.2).
Therefore, we have the following conclusions. Synchrotron mechanisms, as well as IC mechanisms, in general, have incoherent emission.
The jitter mechanism in our scenario is linked to the plasma turbulence that can be one source of the
perturbation, and the radiating electrons are ``bunched'' in a coherent length scale. Thus, the jitter radiation can be viewed as a kind of
collective plasma radiation process \citep{melrose91,melrose17}, and it is therefore the coherent emission. We further stress this point below.

The length scale of jitter radiation can be at sub-Larmor radius, and an electron with each deflection may have a large deflection
angle \citep{medvedev11}. However, the gross deflection angle by the multiple deflections may still be within the radiation cone that is consistent
with the line of sight. \citet{fleishman06} described that large deflection only takes effect on the low-frequency radiation.
Statistically, the gross deflection angle $\theta_c$ can be presented by $\theta^2_c=\theta^2_0N$, where $N$ indicates the times of the deflection, and $\theta_0$
is each deflection angle. The gross deflection length scale $l_c$ is presented by $l_c=Nl_0$, where $l_0$ is each deflection length scale.
The radiation frequencies are $\omega_0=c/l_0$ and $\omega_c=c/l_c$, and we have $\theta_0=\omega_0(l_0/c)$.
Jitter radiation requires that the gross deflection is within the electron radiation cone as
$\theta_c<1/\gamma$. If we take the Larmor frequency $\omega_L$ as $\omega_0$ and we take the observational frequency $\omega$ as $\omega_c$,
we obtain $\omega>\gamma^2\omega_L$. 
The condition can be satisfied because we consider the emission in the TeV energy band in this paper.
This condition is consistent with the result obtained in Section 2.3.2.    

The jitter radiation for the application of GRB emission beyond the GeV energy band was initially suggested by \citet{mao11}.
The spectral properties of the GRBs detected by the Fermi satellite were consistent with the spectral analysis of the jitter radiation \citep{mao20}.
We further explored that the jitter radiation has the possibility to reach the TeV energy band in GRB cases \citep{mao21}.
We examine our scenario and apply the model to the TeV emission of GRB 221009A in the following subsections.

\subsection{Model Description}
The jitter radiation to interpret the GRB emission in the high-energy bands in our model was comprehensively
described in \citet{mao11} and \citet{mao21}. We present the jitter flux of a single electron, which is written as
$I_\omega=\frac{e^4}{m^2c^3\gamma^2}\int_{1/2\gamma^2_\ast}^{\infty}d(\frac{\omega'}{\omega})(\frac{\omega}{\omega'})^2(1-\frac{\omega}{\omega'\gamma_\ast^2}+\frac{\omega^2}{2\omega'^2\gamma_\ast^4})\int dq_0dq\delta(\omega'-q_0+qc)K(q)\delta[q_o-q_o(q)]$, where $\gamma^{-2}_\ast=(\gamma^{-2}+\omega^2_{pe}/\omega^2)$, $\omega'=(\omega/2)(\gamma^{-2}+\theta^2+\omega^{2}_{pe}/\omega^2)$, and $\omega_{pe}$ is the plasma frequency. Here, the dispersion relation $q_0=cq[1\pm\sqrt{1+4\omega_{pe}/c^2q^2\gamma^2}]^{1/2}$
is in the perturbation field, and the radiation $\omega'$ is linked to the perturbation as $\omega'=q_0-qc$. We have the term $B^2=K(q)\sim \int_{q_\nu}^{\eta} q'^{-\zeta_p}dq'$, and $\zeta_p$ is the
spectral index of the turbulence in our model. The radiation frequency can be estimated as $\omega=\gamma^2cq$, if we assume $\omega>>\omega_{pe}$.
The Prandtl number $Pr=10^{-5}(T^4/n)$ constrains the minimum and maximum
number of wavenumber $q$ as $Pr^{1/2}=q_\eta/q_\nu$, where $T$ is the plasma temperature, and $n$ is the total plasma density. Thus, we have $q_{min}<q<q_{max}$, $q_{min}=q_{\nu}$, and $q_{max}=q_{\eta}$. The viscous scale
$q_\nu=2\pi(R/\Gamma\gamma_t)^{-1}$ is dominated by the viscous eddy of turbulence, where $\Gamma$ is the bulk Lorentz factor of the GRB shock, and $\gamma_t$ is the Lorentz factor
of the relativistic turbulence. Then, we can obtain the resistive scale $q_\eta$.
The jitter radiation flux is $I_\omega\sim \omega^{-(\zeta_p-1)}$. Both the power-law index of the jitter radiation and the
radiation frequency are dependent on the parameter of $\zeta_p$, which is the index of the turbulence spectrum.
Different turbulence spectra are determined by turbulence cascade (see a review given by Alexakis \& Biferale 2018).
In particular, the progress on the spectral properties of the kinetic turbulence has been obtained (e.g., Comisso \& Sironi 2022;
Pezzi et al. 2022; Vega et al. 2022). In this paper, we simply take the range of $\zeta_p$ from 5/3 to 2.8, as 5/3 is the typical
Kolmogorov spectral index and 2.8 is the spectral index when turbulence is terminated at the electron damping scale \citep{howes15}.

The jitter radiation frequency has a range of
\begin{equation}
\gamma^2_{min}cq_\nu<\omega<\gamma^2_{max}cq_\eta. 
\end{equation}
It is expected that the jitter radiation of a single electron in our model has a wide frequency range. To be compared, it is well known that synchrotron radiation of a single electron
has a narrow frequency range.
The lower limit of the jitter radiation frequency is $\omega_{min}=\gamma_{min}^2cq_\nu$. Here, we take $\gamma_{min}=1.0$, and we obtain
\begin{equation}
\omega_{min}=18.9(\frac{\gamma_{min}}{1.0})^2(\frac{R}{1.0\times 10^{13}\rm{cm}})^{-1}(\frac{\Gamma}{100.0})(\frac{\gamma_t}{10.0})~\rm{Hz}.  
\end{equation}
The jitter self-absorption should be considered when the jitter radiation occurs at low frequencies \citep{workman08,mao17}. 
The maximum jitter radiation frequency will be given in the following subsections.

\subsection{Constraints on Maximum Radiation Frequency}


\subsubsection{Electron Acceleration Limited by Radiation}
Relativistic electrons obtaining acceleration have radiation cooling when they are moving in a magnetic field.
Here, we should consider the electron cooling effect, and the maximum electron Lorentz factor should be limited by the
electron radiation.
We select a suitable acceleration mechanism (e.g., Honda \& Honda 2005; Aardaneh et al. 2015; Lemoine 2015; Zhdankin et al. 2017; Warren et al. 2021; Comisso \& Sironi 2022; Groselj et al. 2022) that can be coupled with the jitter radiation scenario in this paper.
It is considered that turbulence appears behind shock propagation. Particle diffusion by the filamentary
turbulence occurs \citep{honda05}.
The acceleration timescale can be calculated as
$t_{acc}=(6\pi/8)(c/R)(\gamma m_ec^2/eBU)^2$, where $U\sim c$ is the shock upstream speed, and R is the total length scale of the acceleration region \citep{honda05,mao11}.
We take $R=1.0\times 10^{13}$ cm as the length scale of the system. 
The cooling timescale of the jitter radiation is $t_{cool}=6\pi m_ec/\sigma_T\gamma B^2$,
where $\sigma_T=6.65\times 10^{-25}\rm{cm^2}$ is the Thomson scattering constant.
When we take $t_{acc}=t_{cool}$, the electron Lorentz factor is presented by
\begin{equation}
  \gamma=(\frac{8\sqrt{6}eR}{m_ec^3\sigma_T})^{1/3}=1.8\times 10^8(\frac{R}{1.0\times 10^{13}\rm{cm}})^{1/3}.
\end{equation}
Here, we note that the electron Lorentz factor is not related to the magnetic field.
The number is roughly consistent with that calculated from the shock acceleration in a microturbulent magnetic field \citep{warren21}.
We can use the electron Lorentz factor calculated above to re-examine the maximum jitter radiation frequency as
\begin{equation}
\omega_{max}=1.1\times 10^4(\frac{\gamma}{1.8\times 10^8})^2(\frac{n}{1.0\times 10^{10}\rm{cm^{-3}}})^{-1/2}(\frac{T}{1.2\times 10^{10}\rm{K}})^2(\frac{R}{1.0\times 10^{13}\rm{cm}})^{-1}(\frac{\Gamma}{100.0})(\frac{\gamma_t}{10.0})\rm{TeV}.
\end{equation}
It seems that the jitter radiation in the GRB case can reach the TeV energy band. 

\subsubsection{Turbulent Cascades Limited by Landau Damping}
 When we consider the acceleration by the kinetic turbulence, the maximum electron Lorentz factor can be obtained within the
turbulence length scale. Therefore,
the maximum electron Lorentz factor is $\gamma_{max}=eB/q_{\nu}m_ec^2=eq_\nu^{-(\zeta_p+1)/2}/m_ec^2\sqrt{\zeta_p-1}$, where the magnetic field generated by the turbulence
is $B=q_\nu^{(1-\zeta_p)/2}/\sqrt{\zeta_p-1}$.
We obtain
\begin{equation}
\gamma_{max}=\gamma_0(\frac{R}{1.0\times 10^{13}\rm{cm}})^{(\zeta_p+1)/2}(\frac{\Gamma}{100.0})^{-(\zeta_p+1)/2}(\frac{\gamma_t}{10.0})^{-(\zeta_p+1)/2}
\end{equation}
where $\gamma_0=(6.3\times 10^{10})^{-(\zeta_p+1)/2}e/m_ec^2=5.9\times 10^{-4}(6.3\times 10^{10})^{-(\zeta_p+1)/2}$.
When we take the number of $\zeta_p$ from $=5/3$ to 2.8, the maximum electron Lorentz factor can reach a number from $1.1\times 10^9$ to $1.7\times 10^{14}$.

Turbulence has been comprehensively investigated, and turbulent cascades can be damped at different length scales.
The kinetic turbulence may be terminated by the Landau damping. 
The electron Landau damping of the kinetic turbulence provides \citep{sch09,howes15,comisso22}
\begin{equation}
q_{\eta}r_L\sim 1.
\end{equation}  
Thus, the maximum jitter frequency is limited as $\omega_{max}=\gamma^2_{max}c/r_L=ceB\gamma_{max}/m_ec^2$.
When we take the electron Lorentz factor described by Equation (5) and we obtain $\omega_{max}=(e^2/m_e^2c^3)(q_\nu^{-\zeta_p}/\zeta_p-1)$.
We rewrite it as
\begin{equation}
\omega_{max}=\omega_0(\frac{R}{1.0\times 10^{13}\rm{cm}})^{\zeta_p}(\frac{\Gamma}{100.0})^{-\zeta_p}(\frac{\gamma_t}{10.0})^{-\zeta_p}~\rm{Hz},  
\end{equation}  
where $\omega_0=1.0\times 10^4(6.3\times 10^{-10})^{-\zeta_p}/(\zeta_p-1)$. 
Within the range of $\zeta_p$ from 5/3 to 2.8, we obtain the range of the maximum jitter frequency from 135.0 keV to
$1.3\times 10^3$ TeV. We note that $\zeta_p$ at the Landau damping scale is 2.8. 
It is indicated that GRB emission at the high-energy bands can be explained by the jitter radiation in our scenario even if we consider the turbulence termination by the Landau damping.

\subsection{TeV Emission of GRB 221009A}
It has been reported that GRB 221009A was detected at 18 TeV by LHAASO. 
It seems that the frequency of the jitter radiation in our scenario can reach the TeV energy band.
We then consider the jitter radiation flux in the TeV energy band to compare with the TeV emission from GRB 221009A.
The jitter radiation of a single electron is described as $I_\omega$ in Section 2.2.
The radiation is considered within the range of 1.0 to 20.0 TeV.
We assume the electron energy distribution to produce the TeV emission as a power law of $dN(\gamma)/d\gamma=A\gamma^{-p}$,
where $p=2.2$ is the power-law index, and $A$ is the normalized parameter.
The radiation volume is taken as $V=4\pi R^2H$, and the thickness of the fireball is $H\sim R/\Gamma^2$.
We take $R=1.0\times 10^{13}$ cm and $\Gamma=100.0$. Thus, we can calculate the total numbers of the electrons $N_e$ to have TeV emission in the volume. 
The total flux should be $N_e^2I_\omega$ because the jitter radiation in our scenario is assumed to be coherent. 
The redshift of 0.151 is given to calculate the distance from the source to the Earth.
In the meanwhile, the effect of the strong absorption of the extragalactic background light (EBL) should be considered.
For example, when we take the EBL models of \citet{fran08} and \citet{dom11}, the intrinsic flux can be reduced by an order of $10^{-8}-10^{-9}$ \citep{sahu22}. Here, we take the minimum order of $1.0\times 10^{-9}$ in our calculation. We note that the EBL model from \citet{finke10} can only reduce the intrinsic flux by an order of about $10^{-5}$. 
Our result is strongly dependent on the parameters of $\gamma$, $\zeta_p$, and $A$.
For example, if we have the parameters of $\gamma=1.0\times 10^7$ and $A=3.0\times 10^{-5}~\rm{cm^{-3}}$, we obtain the flux 
to be $1.2\times 10^{-7} ~\rm{erg~cm^{-2}~s^{-1}}$ for $\zeta_p=5/3$ and $7.0\times 10^{-11} ~\rm{erg~cm^{-2}~s^{-1}}$ for $\zeta_p=2.8$.
We put the results in Figure 1.
When we consider the Landau damping effect from Equation (7), the number of $\zeta_p$ is 2.59 if the radiation frequency is at 18 TeV. We keep the parameters of $\gamma=1.0\times 10^7$ and $A=3.0\times 10^{-5}~\rm{cm^{-3}}$, then we obtain the jitter radiation flux of $2.5\times 10^{-10}~\rm{erg~cm^{-2}~s^{-1}}$. We may put this number as the lower limit of our prediction. We further check that the radiation frequency obtained by Equation (4) in the condition
of $\gamma=1.0\times 10^{-7}$ is 34 TeV. Thus, we think that at least some electrons with $\gamma=1.0\times 10^7$ have effective cooling by the jitter radiation.    
The theoretical results with the parameters of $\zeta_p$ and $A$ from our scenario can be constrained by the observational
results of GRB 221009A in the TeV energy band.


\section{Discussion}
Jitter radiation is incoherent emission in general \citep{medvedev00}. However, in this paper, we suggest that the jitter radiation in our model
is coherent. We provide two reasons. First, kinetic turbulence provides a perturbation field to jitter
radiation in our scenario, and the jitter radiation can have emission in a small coherent length scale. Second, the electrons emitting very high-energy photons have the jitter radiation cone along the line of sight, and the jitter radiation is in the coherent length scale. 
Thus, we think that the jitter radiation presented in this paper is coherent emission.
We expect that jitter radiation can have wide application to relativistic jets, and it was already mentioned by \citet{nishikawa05}.

The physical condition of the jitter radiation as a coherent emission can be further explored. \citet{ioka05} suggested coherent radiation
  in relativistic shocks. A current filament has a radius that is approximated to be the plasma skin depth of $\lambda\sim 33.0$ cm if the electron number density is $1.0\times 10^{10}~\rm{cm^{-3}}$. The filament is bent with a curvature radius $r$, and $r\ge r_L$. \citet{ioka05} considered the case that
  a relativistic electron has an emission cone of $1/\gamma$. It provides the case of $r> \gamma^2\lambda$. As presented in Section 2.1, this is the condition that we have applied in this paper. In this scenario, all the electrons are ``bunched'' in a coherent length scale. \citet{ioka05} estimated that the ratio between the coherent cooling length scale and the radius $r$ is about 0.15 and this constant is not dependent on other parameters. It means that the electron cooling has a very short cooling timescale in a very short cooling distance. In principle, this result is consistent with the estimation of the cooling timescale given by the former work from \citet{mao11}. On the other hand, \citet{ioka05} suggested another condition of $\gamma^2\lambda\ge r>\lambda$. It is indicated that some electrons are not ``bunched'' and the emission of these electrons outside the filament is not coherent. In such cases, the jitter radiation as a coherent emission is invalid.   

The EBL absorption takes effect on the gamma-ray photons beyond about 1 TeV. GRB 221009A with the emission at 18 TeV has a redshift of 0.151 \citep{deugarte22}. \citet{zhang22} suggested a proton synchrotron model to produce the TeV emission of GRB 221009A, and a reasonable EBL
absorption was considered. Thus, the new physics (e.g., Lorentz invariance violation or axion-like particle) is not necessary.  
In order to further examine our scenario, we hope that the jitter radiation in our model can be applied to some kinds of high-energy sources in the Galaxy in order to avoid the strong EBL effect that has large uncertainty. 

Kinetic turbulence takes the same acceleration mechanism for both electrons and protons \citep{comisso22}. 
In our model, we can assume that protons have the same acceleration process as electrons. Thus, from Equation (5), we obtain that
the protons have the Lorentz factor of $6.0\times 10^{5}-9.3\times 10^{10}$. In other words, protons can be accelerated to be
from 0.2 PeV to 31 EeV. It indicates that GRB might be one kind of the ultrahigh-energy cosmic-ray sources.

\section{Conclusion}
We utilize the jitter radiation in our scenario to explain the GRB 221009A emission at 18 TeV.
In this paper, we illustrate that the jitter radiation can reach the TeV energy band in the GRB case.
We take the jitter radiation as a kind of coherent emission. In our scenario, we suggest that GRBs are both extremely high-energy sources and high-energy
cosmic-ray sources.
Our modeling results can be effectively constrained by the observation of GRB 221009A in the TeV energy band.  
We expect that the jitter radiation in our scenario has wide application in the research field of the high-energy astrophysics.


\acknowledgments
J.M. is supported by the National Natural Science Foundation of China 11673062 and the Oversea Talent Program of Yunnan Province.

\begin{figure}
\includegraphics[width=\textwidth]{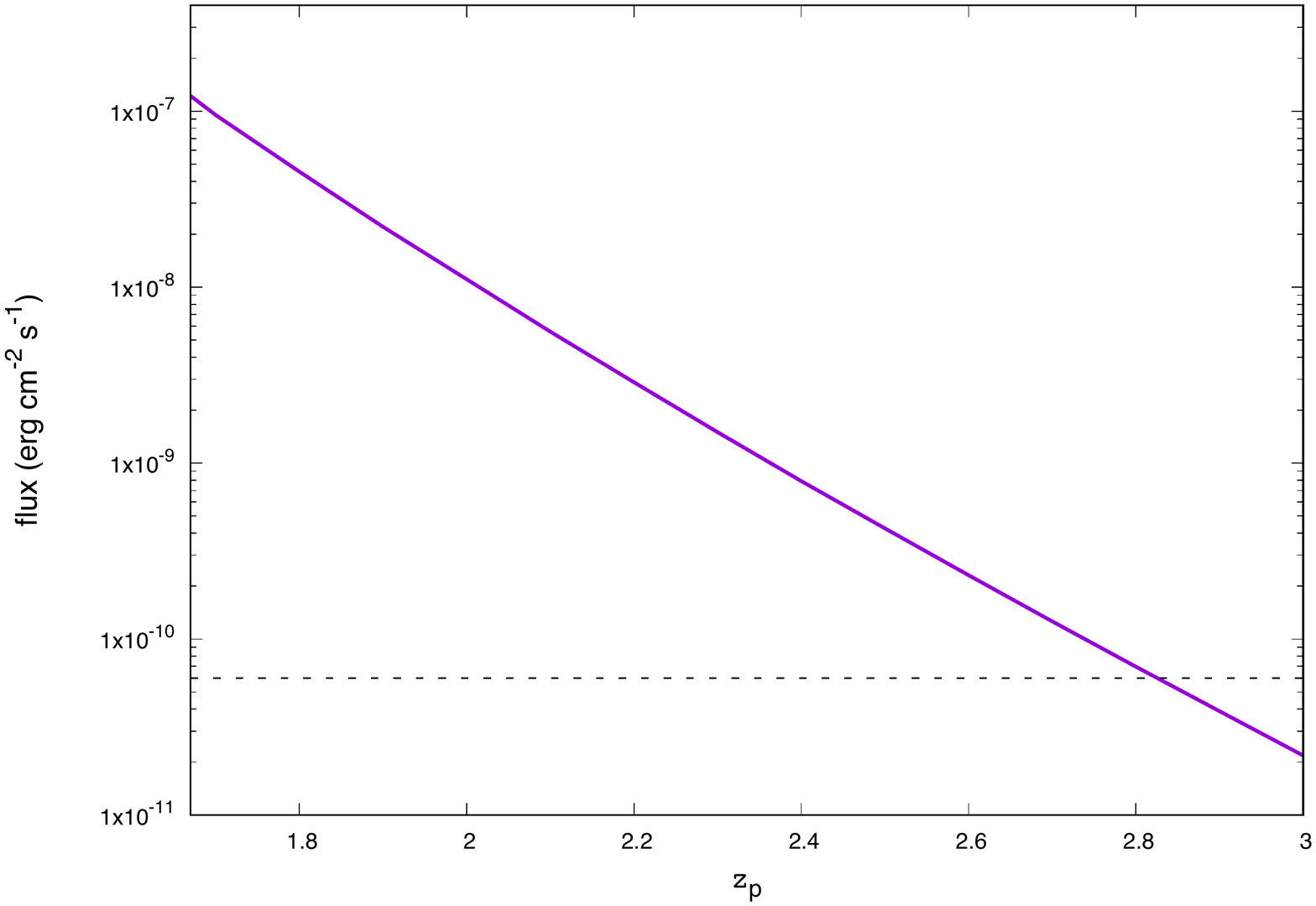}
\caption{The jitter radiation flux as a function of the spectral index $\zeta_p$ of the turbulence. The solid line (purple) indicates the calculation result with the parameters of $\gamma=1.0\times 10^7$ and $A=3.0\times 10^{-5}~\rm{cm^{-3}}$. The dashed line (grey) is the LHAASO sensitivity (2000 s) at 18 TeV \citep{cao19,ren22,sahu22}.
  \label{fig1}}
\end{figure}


\begin{thebibliography}{}
\bibitem[Abdalla et al.(2019)]{abdalla19} Abdalla, H., Adam, R., Aharonian, F., et al. 2019, Nature, 575, 464
\bibitem[Alexakis \& Biferale(2018)]{alexakis18} Alexakis, A., \& Biferale, L. 2018, Physics Reports, 767, 1    
\bibitem[Ardaneh et al.(2015)]{ardaneh15} Ardaneh, K., Cai, D., Nishikawa, K.-I., \& Lembege, B. 2015, \apj, 811, 57
\bibitem[Cao et al.(2022)]{cao19} Cao, Z., della Volpe, D., Liu, S., et al. 2022, Chinese Physics C, 46, 035001
\bibitem[Comisso \& Sironi(2022)]{comisso22} Comisso, L., \& Sironi, L. 2022, ApJL, 936, L27
\bibitem[Daigne et al.(2011)]{daigne11} Daigne, F., Bosnjak, Z., \& Dubus, G. 2011, \aap, 526, A110
\bibitem[Dominguez et al.(2011)]{dom11} Dominguez, A., Primack, J. R., Rosario, D. J., et al. 2011, \mnras, 410, 2556
\bibitem[de Ugarte Postigo et al.(2022)]{deugarte22} de Ugarte Postigo, A., Izzo, L., Pugliese, G., et al. 2022, GCN circ. 32648 
\bibitem[Derishev \& Piran(2021)]{derishev21} Derishev, E., \& Piran, T. 2021, \mnras, 923, 135
\bibitem[Finke et al.(2010)]{finke10} Finke, J. D., Razzaque, S. \& Dermer, C. D. 2010, \apj, 712, 238 
\bibitem[Fleishman(2006)]{fleishman06} Fleishman, G. D. 2006, \apj, 638, 348  
\bibitem[Fraija et al.(2020)]{fraija20} Fraija, N., Laskar, T., Dichiara, S., Beniamini, P., Duran, R., B., Dainotti, M. G., \& Becerra, R. L. 2020, \apj, 905, 112
\bibitem[Franceschini et al.(2008)]{fran08} Franceschini, A., Rodighiero, G., \& Vaccari, M. 2008, \aap, 487, 837
\bibitem[Ghisellini et al.(2020)]{ghisellini20} Ghisellini, G., Ghirlanda, G., Oganesyan, G., et al. 2020, \aap, 636, A82 
\bibitem[Gill et al.(2020)]{gill20} Gill, R., Granot, J., \& Beniamini, P. 2020, \mnras, 499, 1356
\bibitem[Gonzalez et al.(2022)]{gonzalez22} Gonzalez, M. M., Avila Rojas, D., Pratts, A., Hernandez-Cadena, S., Fraija, N., Alfaro, R., Perez Araujo, Y., \& Montes, J. A. 2023, ApJ, 944, 178 
\bibitem[Groselj et al.(2022)]{groselj22} Groselj, D., Sironi, L., \& Beloborodov, A. M. 2022, \apj, 933, 74
\bibitem[H.E.S.S. collaboration(2021)]{hess21} H.E.S.S. collaboration, 2021, Sci, 372, 1081
\bibitem[Howes(2015)]{howes15} Howes, G. G. 2015, Astrophysics and Space Science Libray, 407, 123
\bibitem[Honda \& Honda(2005)]{honda05} Honda, M., \& Honda, Y. S. 2005, \apj, 633, 733
\bibitem[Huang et al.(2022)]{lhaaso22} Huang, Y., Hu, S., Chen, S., et al. 2022, GCN circ. 32677 
\bibitem[Ioka(2006)]{ioka05} Ioka, K. 2006, Prog. Theor. Phys., 114, 1317
\bibitem[Kelner et al.(2013)]{kelner13} Kelner, S. R., Aharonian, F. A., \& Khangulyan, D. 2013, \apj, 774, 61 
\bibitem[Lemoine(2015)]{lemoine15} Lemoine, M. 2015, \apj, Journal of Plasma Physics, 81, 455810101
\bibitem[MAGIC Collaboration(2019a)]{magic19a} MAGIC collaboration, 2019a, Nature, 575, 455
\bibitem[MAGIC Collaboration(2019b)]{magic19b} MAGIC collaboration, 2019b, Nature, 575, 459  
\bibitem[Mao \& Wang(2011)]{mao11} Mao, J., \& Wang, J. 2011, \apj, 731, 26
\bibitem[Mao \& Wang(2017)]{mao17} Mao, J., \& Wang, J. 2017, \apj, 838, 78
\bibitem[Mao et al.(2020)]{mao20} Mao, J., Li, L., \& Wang, J. 2020, \apj, 898, 14
\bibitem[Mao \& Wang(2021)]{mao21} Mao, J., \& Wang, J. 2021, \mnras, 505, 4608   
\bibitem[Medvedev \& Loeb(1999)]{medvedev99} Medvedev, M. V., \& Loeb, A. 1999, \apj, 526, 697
\bibitem[Medvedev(2000)]{medvedev00} Medvedev, M. V. 2000, \apj, 540, 704 
\bibitem[Medvedev(2006)]{medvedev06} Medvedev, M. V. 2006, \apj, 637, 869
\bibitem[Medvedev et al.(2011)]{medvedev11} Medvedev, M. V., Frederiksen, J. T., Haugbolle, T., \& Nordlund, A. 2011, \apj, 737, 55   
\bibitem[Melrose(1991)]{melrose91} Melrose, D. B. 1991, ARA\&A, 29, 31
\bibitem[Melrose(2017)]{melrose17} Melrose, D. B. 2017, Reviews of Modern Plasma Physics, 1, 5
\bibitem[Nishikawa et al.(2005)]{nishikawa05} Nishikawa, K.-I., Hardee, P., Richardson, G., Preece, R., Sol, H., \& Fishman, G. J. 2005, \apj, 622, 927 
\bibitem[Omodei et al.(2022)]{omodei22} Omodei, N., Bruel, P., Bregeon, J., Pesce-Rollins, M., Horan, D., Bissaldi, E., \& Pillera, R. 2022, GCN circ. 32760
\bibitem[Pezzi et al.(2022)]{pezzi22} Pezzi, O., Blasi, P., \& Matthaeus, W. H. 2022, \apj, 928, 25
\bibitem[Ren et al.(2022)]{ren22} Ren, J., Wang, Y., \& Zhang, L.-L., 2022, arXiv: 2210.10673
\bibitem[Sahu et al.(2023)]{sahu22} Sahu, S., Medina-Carrillo, B., Sanchez-Colon, G., \& Subhash, R. 2023, ApJ, 942, L30
\bibitem[Schekochihin et al.(2009)]{sch09} Schekochihin, A. A., Cowley, S. C., Dorland, W., Hammett, G. W., Howes, G. G., Quataert, E., \& Tatsuno, T.
  2009, ApJS, 182, 310
\bibitem[Vega et al.(2022)]{vega22} Vega, C., Boldyrev, S., Roytershteyn, V., \& Medvedev, M. 2022, ApJL, 924, L19 
\bibitem[Warren et al.(2021)]{warren21} Warren, D. C., Beauchemin, C. A. A., Barkov, M. V., \& Nagataki, S. 2021, \apj, 906, 33   
\bibitem[Workman et al.(2008)]{workman08} Workman, J. C., Morsony, B. J., Lazzati, D., \& Medvedev, M. V. 2008, \mnras, 386, 199
\bibitem[Zhang et al.(2022)]{zhang22} Zhang, B. T., Murase, K., Ioka, K., Song, D., Yuan, C., \& Meszaros, P. 2022, arXiv: 2211.05754   
\bibitem[Zhdankin et al.(2017)]{zhdankin17} Zhdankin, V., Werner, G. R., Uzdensky, D. A., \& Begelman, M. C. 2017, Phys. Rev. Lett., 118, 055103
\end{thebibliography}
\end{document}